\documentclass[aps,prl,showpacs,amsmath,amssymb,amsfonts,superscriptaddress,twocolumn,lengthcheck]{revtex4-2}

\usepackage{ulem}
\usepackage{color}
\usepackage{ulem}
\usepackage{amsmath}
\usepackage{amssymb}
\usepackage{graphics}
\usepackage{graphicx}
\usepackage{dcolumn}
\usepackage[pdfpagemode=UseNone,pdfstartview=FitH,colorlinks=true,linkcolor=blue,urlcolor=blue,anchorcolor=blue,citecolor=blue]{hyperref}
\usepackage{lineno}
\usepackage{booktabs}

\begin{document}
\title{Macroscopic Bell state between a millimeter-sized spin system and a superconducting qubit}

\author{Da Xu}
\thanks{These two authors contributed equally}
\affiliation{Zhejiang Key Laboratory of Micro-Nano Quantum Chips and Quantum Control, School of Physics, and State Key Laboratory for Extreme Photonics and Instrumentation, Zhejiang University, Hangzhou 310027, China}

\author{Xu-Ke Gu}
\thanks{These two authors contributed equally}
\affiliation{Zhejiang Key Laboratory of Micro-Nano Quantum Chips and Quantum Control, School of Physics, and State Key Laboratory for Extreme Photonics and Instrumentation, Zhejiang University, Hangzhou 310027, China}

\author{Yuan-Chao Weng}
\affiliation{Zhejiang Key Laboratory of Micro-Nano Quantum Chips and Quantum Control, School of Physics, and State Key Laboratory for Extreme Photonics and Instrumentation, Zhejiang University, Hangzhou 310027, China}

\author{He-Kang Li}
\affiliation{Zhejiang Key Laboratory of Micro-Nano Quantum Chips and Quantum Control, School of Physics, and State Key Laboratory for Extreme Photonics and Instrumentation, Zhejiang University, Hangzhou 310027, China}

\author{Yi-Pu Wang}
\email[Corresponding author.~Email:~]{ypwang@zju.edu.cn}
\affiliation{Zhejiang Key Laboratory of Micro-Nano Quantum Chips and Quantum Control, School of Physics, and State Key Laboratory for Extreme Photonics and Instrumentation, Zhejiang University, Hangzhou 310027, China}

\author{Shi-Yao Zhu}
\affiliation{Zhejiang Key Laboratory of Micro-Nano Quantum Chips and Quantum Control, School of Physics, and State Key Laboratory for Extreme Photonics and Instrumentation, Zhejiang University, Hangzhou 310027, China}
\affiliation{College of Optical Science and Engineering, Zhejiang University, Hangzhou 310027, China}	
\affiliation{Hefei National Laboratory, Hefei 230088, China}

\author{J. Q. You}
\email[Corresponding author.~Email:~]{jqyou@zju.edu.cn}
\affiliation{Zhejiang Key Laboratory of Micro-Nano Quantum Chips and Quantum Control, School of Physics, and State Key Laboratory for Extreme Photonics and Instrumentation, Zhejiang University, Hangzhou 310027, China}
\affiliation{College of Optical Science and Engineering, Zhejiang University, Hangzhou 310027, China}	

\date{\today}
\begin{abstract}
Entanglement is a fundamental property in quantum mechanics that systems share inseparable quantum correlation regardless of their mutual distances. Owing to the fundamental significance and versatile applications, the generation of quantum entanglement between {\it macroscopic} systems has been a focus of current research. Here we report on the deterministic generation and tomography of the macroscopically entangled Bell state in a hybrid quantum system containing a millimeter-sized spin system ($\sim 1\times10^{19}$ atoms) and a micrometer-sized superconducting qubit. The deterministic generation is realized by coupling the macroscopic spin system and the qubit via a microwave cavity. Also, we develop a joint tomography approach to confirming the deterministic generation of the Bell state, which gives a generation fidelity of $0.90\pm0.01$. Our work makes the macroscopic spin system the {\it largest} system (in the sense of atom number) capable of generating the maximally entangled quantum state.
\end{abstract}


\maketitle
\section{INTRODUCTION}
Hybrid quantum systems harness the advantages of different subsystems to implement quantum information processing and other quantum technology applications~\cite{Xiang-RMP-13,Schmiedmayer-PNAS-15}. Recently, the hybrid quantum system based on collective spin excitations in ferromagnetic materials becomes a promising platform for quantum information and quantum engineering~\cite{Lachance-APE-2019,Yuan-PR-22,Blanter-PR-22}, especially for the quantum transducer applications in a quantum network~\cite{Kimble-Nature-2008,Li-PRXQuantum-2021,Hisatomi-PRB-2016,Osada-PRL-2016,Zhang-PRL-2016,Haigh-PRL-2016}. This is because the quantum of these collective excitations (magnon) is capable of coupling many different systems, including  optical photons~\cite{Hisatomi-PRB-2016,Osada-PRL-2016,Zhang-PRL-2016,Haigh-PRL-2016}, microwave photons~\cite{Huebl-PRL-2013,Tabuchi-PRL-2013,Zhang-PRL-2014,Tobar-PRApp-2014,Hu-PRL-2015,You-npj-2015},  phonons~\cite{Zhang-SA-2016,Potts-PRX-2021,Shen-PRL-2022,Kheirabady-JPB-2023}, and superconducting qubits~\cite{Tabuchi-Science-2015,Nakamura-PRL-2017,Quirion-SA-2017,Lachance-quirion-Science-2020,Xu-PRL-2023}. More importantly, the large size of the ferromagnetic spin system ($\sim 1$~mm) and the enormous number of spins in it ($\sim 10^{19}$) also make it an ideal platform for testing some fundamental properties in quantum mechanics, such as the quantum entanglement between macroscopic systems~\cite{Julsgaard-Nature-2001,Klimov-SA-2015,Lee-Science-2011,Riedinger-Nature-2018,Agarwal-PRResearch-2019,Thomas-NP-2021,Kotler-Science-2021,
Chu-Science-2023,Fink-Science-2023}. Indeed, quantum entanglement between macroscopic objects containing a very large number of atoms is of particular importance and has been demonstrated in various systems, such as the circuit QED system~\cite{Steffen-Science-2006} and the mechanical resonator~\cite{Wollack-Nature-2022}. In the hybrid magnonic system, entanglement has been used as the resource to detect the magnon number~\cite{Lachance-quirion-Science-2020}, where the entanglement plays the same role as in the superconducting-qubit dispersive readout process~\cite{Jeffrey-PRL-2014}.
Very recently, we have deterministically generated a single-magnon state and its superposition with the vacuum (zero-magnon state)~\cite{Xu-PRL-2023}. This has removed some barriers towards the generation of an entangled Bell state between the magnonic system and the qubit, which is one of the essential ingredients for quantum transduction. Nevertheless, a convincing characterization of the entangled Bell state by quantum tomography remains the greatest outstanding challenge due to the shorter lifetime of the magnon and the limited number of qubits that can be integrated into the hybrid system.

Here we report the deterministic generation and tomography of the maximally entangled Bell state between a magnonic system and a superconducting qubit. The Bell state is generated in the resonant qubit-magnon coupling regime, using a fast magnon-qubit swap operation. The effective coupling between the magnon and the qubit is mediated by a three-dimensional (3D) microwave cavity~\cite{Tabuchi-Science-2015}, and the qubit frequency is tuned using the dressed Autler-Townes (AT) doublet states~\cite{Xu-PRL-2023}.  To characterize the {\it deterministically} generated magnon-qubit Bell state, we develop a new joint tomography approach. In contrast to the conventional joint tomography technique, which requires separate local measurements of the two subsystems~\cite{Steffen-Science-2006,Linpeng-NJP-2013}, our method only requires to measure the qubit.
Our experiment makes the ferromagnetic spin system the largest system (in the sense of atom number) capable of generating a macroscopic entangled state. It paves a way to use magnonic systems to demonstrate fundamental properties in quantum mechanics and to develop novel quantum devices such as the transducer in a quantum network, where the entangled state is a critical resource~\cite{Lachance-APE-2019}.

\begin{figure}
	\centering
	\includegraphics[width=0.48\textwidth]{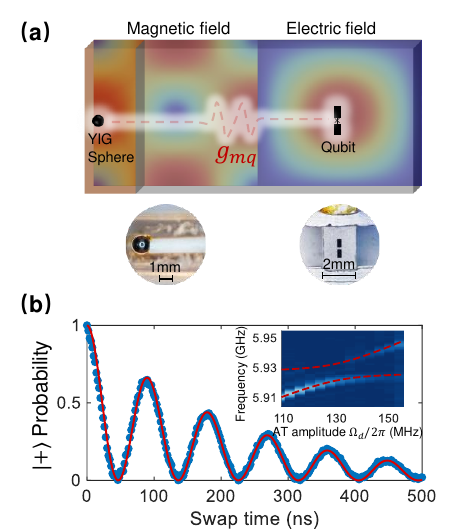}%
	\caption{(a)~Schematic of the hybrid quantum system, where a 1 mm-diameter YIG sphere is placed in the copper part of the 3D cavity, near the magnetic-field antinode of the cavity mode $\rm{TE_{102}}$, and a 3D transmon qubit is mounted in the aluminum part of the 3D cavity, near the electric-field antinode of the cavity mode $\rm{TE_{102}}$. The left (right) half part shows the magnetic (electric) field of the cavity mode $\rm{TE_{102}}$. The optical microscopy images of the YIG sphere and qubit chip are shown below the cavity. (b)~Measured magnon-qubit swap oscillation, where the oscillation frequency $\Omega_{mq}$ gives the magnon-qubit coupling strength via $\Omega_{mq}=2g_{mq}$, i.e., $g_{mq}/2\pi=5.59$~MHz. The red curve corresponds to the numerically simulated results (see Supplementary Materials). Inset: Measured avoided level crossing between the qubit and the magnon. We change the AT drive amplitude (horizontal axis) to tune the qubit frequency while the magnon frequency is fixed.}
	\label{fig1}
\end{figure}

\section{Results}
\subsection{MAGNON-QUBIT STATE SWAP}

The hybrid system in our experiment is composed of a 1~mm-diameter yttrium-iron-garnet (YIG) sphere and a 3D transmon, both of which are placed in a rectangular 3D microwave cavity, cf. Fig.~\ref{fig1}(a) and Methods for details. We consider Kittel-mode magnons in the YIG sphere, with a tunable frequency achieved by changing the strength of the applied magnetic field. Because the magnons and the qubit are dispersively coupled to the cavity mode, their frequencies are slightly modified by the cavity mode (see Sec.~I.A in Supplementary Materials). In the experiment, we fix the cavity-modified magnon frequency at $\omega_m/2\pi=5.927$ GHz. The lowest three eigenstates of the transmon are $\vert g\rangle$, $\vert e\rangle$ and $\vert f\rangle$, with the corresponding cavity-modified transition frequencies from $\vert g\rangle$ to $\vert e\rangle$ and $\vert e\rangle$ to $\vert f\rangle$ being $\omega_{ge}/2\pi=5.847$ GHz and $\omega_{ef}=5.493$ GHz, respectively. In our hybrid system, the tuning speed of the magnon frequency by a magnetic field is slow, so we rely on tuning the transition frequency of the transmon. Conventionally, the transmon can become tunable by replacing the single Josephson junction with a superconducting quantum interference device (SQUID), but its coherence is much reduced by the flux noise due to the strong bias magnetic field applied to the magnons. Thus, we harness the Autler-Townes (AT) effect~\cite{AT-PR-1955,Han-PRApplied-2019} to implement the frequency tunability of the single-junction transmon (see Sec.~I.B in Supplementary Materials). This is achieved by applying a strong control drive (AT drive) in resonance with the $\vert e\rangle$ to $\vert f\rangle$ transition, i.e., $\omega_d=\omega_{ef}$, so as to have both $\vert e\rangle$ and $\vert f\rangle$ dressed to be the doublet states $\vert\pm\rangle=(\vert e\rangle\pm\vert f\rangle)/\sqrt{2}$. The corresponding transition frequencies from $\vert g\rangle$ to the two new excited states $\vert\pm\rangle$ are $\omega_{\pm}=\omega_{ge}\pm\Omega_d/2$, where $\Omega_d$ is the Rabi frequency related to the amplitude of the AT drive. Hereafter, we define $\vert g\rangle$ and $\vert +\rangle$ as the ground and excited states of the qubit, respectively. Thus, we can tune the transition frequency of the qubit $\omega_q\equiv\omega_+$ by changing the amplitude of the AT drive.

The magnon-qubit interaction is mediated by the cavity mode $\rm{TE_{102}}$ with $\omega_{102}/2\pi=6.388$ GHz. Here, both the magnon mode and the qubit are strongly coupled to the cavity mode $\rm{TE_{102}}$. When the magnon mode and the qubit are both far detuned from the cavity mode $\rm{TE_{102}}$, an effective coupling between the magnon mode and the qubit is achieved via exchanging virtual cavity photons~\cite{Imamoglu-PRL-2009}, giving rise to a Jaynes-Cummings Hamiltonian for the magnon-qubit hybrid system (see Sec.~I.C in Supplementary Materials):
\begin{equation}\label{JCH}
	H/\hbar=\frac{1}{2}\omega_q\sigma_z+\omega_ma^\dagger a+g_{mq}(\sigma_+a+\sigma_-a^\dagger),
\end{equation}
where $g_{mq}$ is the effective magnon-qubit coupling strength, $\sigma_z=\vert+\rangle\langle+\vert-\vert g\rangle\langle g\vert$ is a Pauli operator, $\sigma_+=\vert +\rangle\langle g\vert$ and $\sigma_-=\vert g\rangle\langle +\vert$ are ladder operator, and $a$ ($a^\dagger$) is the magnon annihilation (creation) operator. When the magnon mode and the qubit are on resonance, the magnon-qubit interaction enables state swap between them. In Fig.~\ref{fig1}(b), we show the magnon-qubit swapping measured by initializing the qubit in the excited state $\vert +\rangle$ and then tuning the qubit in resonance with the magnon mode for a given period of time. The oscillation frequency $\Omega_{mq}$ is related to the magnon-qubit coupling strength: $\Omega_{mq}=2g_{mq}$, where $g_{mq}/2\pi=5.59$ MHz. In the experiment, the qubit states are measured using the Jaynes-Cummings nonlinearity readout scheme~\cite{Reed-PRL-2010,Xu-PRL-2023} via the cavity mode $\rm{TE_{103}}$ with frequency $\omega_{\rm{TE103}}/2\pi=8.367$ GHz.

\begin{figure}
	\centering
	\includegraphics[width=0.48\textwidth]{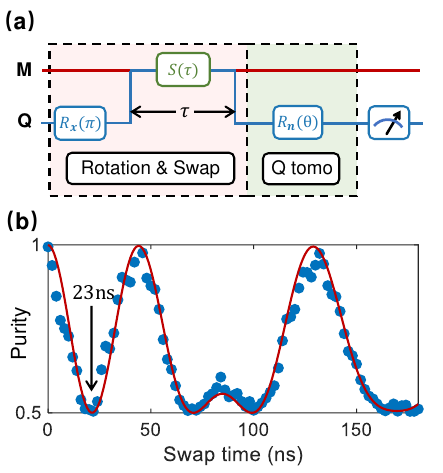}%
	\caption{(a)~Operation sequence for measuring the purity of the qubit subsystem during the magnon-qubit swap process in Fig.~\ref{fig1}(b). Here, `M' denotes the magnon (upper red line) and `Q' denotes the qubit (lower blue line). We first apply a $\pi$ rotation to the qubit and then tune the qubit to be resonant with the magnon for a varying swap time $\tau$. Afterwards, we apply different qubit rotations $R_{\bf n}(\theta)=I, R_x(\pi/2)$, and $R_y(\pi/2)$ to perform state tomography of the qubit (Q tomo). (b)~Purity of the qubit subsystem versus the swap time, obtained from the reconstructed reduced density matrix of the qubit. The dots are the experimental results, and the red curve corresponds to the numerically simulated results (see Supplementary Materials).}
	\label{fig2}
\end{figure}

\subsection{MAGNON-QUBIT BELL STATE}

Below we generate the magnon-qubit Bell state, which is a maximally entangled state. The magnon-qubit system is initialized in the ground state $\vert \psi_{\rm ini}\rangle=\vert 0,g\rangle\equiv\vert 0\rangle\otimes\vert g\rangle$, where $\vert 0\rangle$ and $\vert g\rangle$ are the ground states of the magnon and the qubit, respectively. This ground-state initialization is achieved via energy relaxation, i.e., by keeping the system waiting for 150 $\mu$s after each single-shot measurement. Also, via the AT effect, the transition frequency of the qubit is tuned at $\omega_{q}/2\pi=\omega_{\pi}/2\pi=5.867$~GHz, largely detuned from the magnon frequency $\omega_m$, i.e., $\vert\Delta_{qm}\vert=\vert \omega_q-\omega_m\vert\gg g_{mq}$. Subsequently, the qubit is excited to $\vert +\rangle$ by applying a $\pi$ pulse, and the system is in the state $\vert 0,+\rangle$. Here, $\omega_{\pi}/2\pi$ refers to the ``work point" frequency at which we apply a $\pi$ pulse to the qubit. Finally, the qubit is tuned to be resonant with the magnon for a period of time $\tau=\pi/4g_{qm}\approx23$~ns. The system is then prepared in the magnon-qubit Bell state $\vert\psi_{\rm Bell}\rangle=(\vert 0,+\rangle-i\vert 1,g\rangle)/\sqrt{2}$ (see Sec.~II.A in Supplementary Materials for details about the evolution).

In order to measure the entanglement evolution during the swap process, we harness the purity of the qubit subsystem, which is ${\rm Tr}(\rho_{q}^2)$, where $\rho_{q}$ is the reduced density matrix of the qubit subsystem obtained by implementing the partial trace of the magnon-qubit joint density matrix. Experimentally, we measure the purity by performing a sequence of operations in Fig.~\ref{fig2}(a). The measured purity of the qubit subsystem is shown in Fig.~\ref{fig2}(b), in comparison with the numerically simulated purity. It can be seen that the measured purity is about 0.5 at the swap time $\tau=23$ ns, indicating that the qubit and the magnon really become maximally entangled.

\begin{figure*}
	\includegraphics[width=0.98\textwidth]{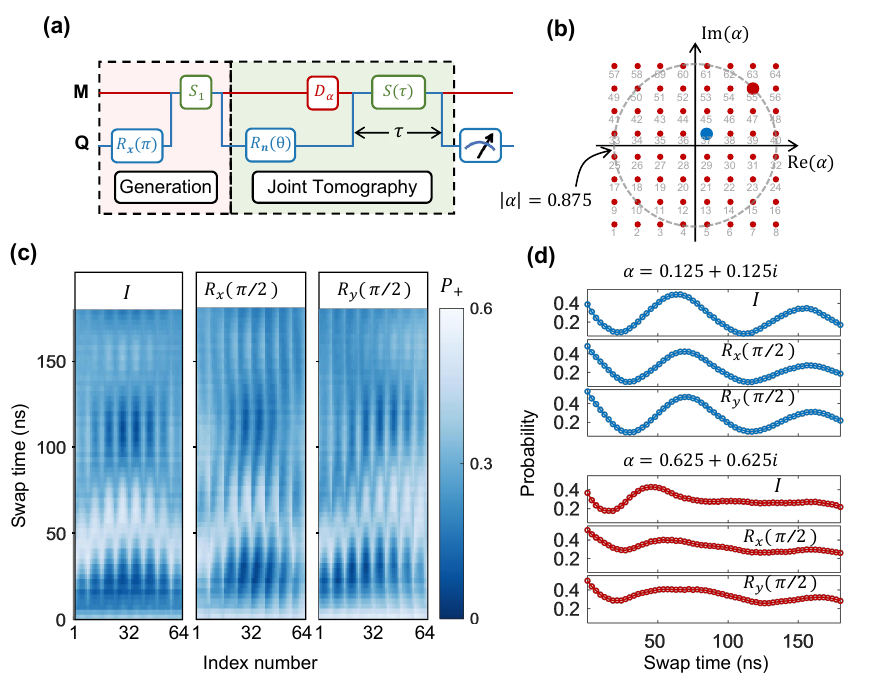}%
	\caption{(a)~Operation sequence for the generation and tomography of the Bell state. Starting from the ground state $\vert 0,g\rangle$ of the magnon-qubit system, we first apply a $R_{x}(\pi)$ rotation ($\pi$ pulse) to the qubit and then tune the qubit to be resonant with the magnon for 23~ns, which is the swap operation $S_1$ . The hybrid magnon-qubit system is prepared in the Bell state $\vert\psi_{\rm Bell}\rangle=(\vert 0,+\rangle-i\vert 1,g\rangle)/\sqrt{2}$. Afterwards, we successively apply three operations on the qubit and the magnon, which are the qubit rotation $R_{\bf n}(\theta)$, the magnon displacement operation $D_\alpha$, and the magnon-qubit swap operation $S(\tau)$. These three operations form the joint state tomography operation $T(R,\alpha,\tau)$. Finally, the qubit is read out.
(b)~The chosen displacement $\alpha$ in the phase space. We choose $8\times8$ different displacements $\alpha$, with the real and imaginary parts ranging from -0.875 to 0.875, and the increment step is 0.25.
(c)~All the $3\times64\times61$ experimental data for reconstructing the density matrix. Every vertical slice is the magnon-qubit swap curve after performing an operation, shown in the sequence of qubit rotations $I$, $R_x(\frac{\pi}{2})$, and $R_y(\frac{\pi}{2})$. The 61 evolution parameters $\tau$ in the swap operation $S(\tau)$ are given in the vertical axis, while other $3\times64$ parameters are given in the horizontal axis. Corresponding to each qubit rotation, the 64 chosen values of $\alpha$ are indicated by index numbers in the horizontal axis. The one-to-one correspondence between these 64 values of $\alpha$ and the index numbers from 1 to 64 are explicitly shown in (b).
(d)~Two set of swap curves for displacement operations with $\alpha=0.125+0.125i$, and $0.625+0.625i$, i.e., the two big dots in (b).}
	\label{fig3}
\end{figure*}

\subsection{JOINT STATE TOMOGRAPHY}

In order to characterize the generated entangled state, we need to perform the joint state tomography of the system~\cite{Steffen-Science-2006,Linpeng-NJP-2013}. Since the magnon and qubit are maximally entangled, the observables measured by the quantum tomography must be composite observables containing the information of both systems. Conventionally, this is implemented by separately measuring the states of the two systems. Then, it requires another ancillary qubit as the detector of the magnon. Here it is difficult because one of the two optimal positions related to the cavity mode ${\rm TE}_{102}$ is occupied by the current qubit, while the other optimal position is close to the position for mounting the YIG sphere, yielding the quantum coherence of the ancillary qubit therein {\it much reduced} by the strong bias magnetic field applied to the YIG sphere. Moreover, the coupling between the magnon and qubit is achieved via the cavity mode ${\rm TE}_{102}$, which can also induce {\it unwanted} coupling between the current qubit and the ancillary qubit. Thus, we develop a new tomography method to reconstruct quantum states of the magnon-qubit system (see Sec.~II.B in Supplementary Materials), {\it without} using an ancillary qubit.

Starting from the generated Bell state $\rho$, we first apply a qubit rotation $R=R_{\bf n}(\theta)$ with rotation axis ${\bf n}$ and angle $\theta$. This is realized by applying a microwave drive (in resonance with the qubit) with a given amplitude and phase. Then, the magnon displacement operation $D_\alpha$ is applied via a microwave drive with a given displacement amplitude and phase. Afterwards, we tune the qubit to be resonant with the magnon for implementing the magnon-qubit swapping $S(\tau)$. Finally, the probability of the qubit's excited state $P_+$ is read out (cf. Ref.~\cite{Xu-PRL-2023} for details about the readout method). These three operations are combined as the {\it joint} tomography operation $T(R,\alpha,\tau)\equiv S(\tau)D_\alpha R$, as shown in Fig.~\ref{fig3}(a). The measurement results are given by (cf. Sec.~II.B in Supplementary Materials)
\begin{equation}
E(R,\alpha,\tau)={\rm Tr}[\rho O(R,\alpha,\tau)],
\end{equation}
with
\begin{equation}
O(R,\alpha,\tau)\equiv T^\dagger(R,\alpha,\tau)\vert +\rangle\langle +\vert T(R,\alpha,\tau).
\end{equation}
The joint tomography operation $T(R,\alpha,\tau)$ effectively changes the observable into {\it composite} observables $O(R,\alpha,\tau)$. Here we use three different qubit rotations: $R=I, R_x(\frac{\pi}{2})$, and $R_y(\frac{\pi}{2})$, $8\times8$ different magnon displacements $\alpha$ [see Fig.~\ref{fig3}(b)], and 61 different swap times $\tau$ (0 to 180~ns), which form a set of composite observables $\{O(R,\alpha,\tau)\}$ to provide the needed information of the generated state $\rho$ prior to the joint tomography operation. All these $3\times64\times61$ swap data are shown in Fig.~\ref{fig3}(c). For clarity, two typical set of swap curves with $\alpha=0.125+0.125i$ and $0.625+0.625i$ are shown in Fig.~\ref{fig3}(d), where the three curves in each set correspond to the qubit rotations $R=I, R_x(\frac{\pi}{2})$, and $R_y(\frac{\pi}{2})$, respectively.

\begin{figure}
	\includegraphics[width=0.49	\textwidth]{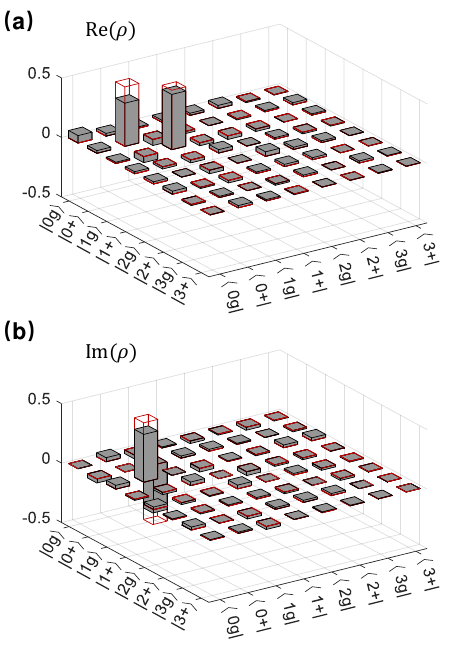}%
	\caption{(a)~Real part of the density matrix $\rho$ for the reconstructed Bell state. The red boxes correspond to the real part of the density matrix of the ideal Bell state $\vert\psi_{\rm Bell}\rangle=(\vert 0,+\rangle-i\vert 1,g\rangle)/\sqrt{2}$. (b)~Imaginary part of the density matrix $\rho$ for the reconstructed Bell state. The red boxes correspond to the imaginary part of the density matrix of the ideal Bell state $\vert\psi_{\rm Bell}\rangle$. The number of magnons in the Fock states is truncated at $n=3$ for clarity. }
	\label{fig4}
\end{figure}

With the measured expectation values $E(R,\alpha,\tau)$ for $3\times 64\times 61$ different observables $O(R,\alpha,\tau)$, we can obtain the most probable state $\rho$ using the convex optimization, i.e., by minimizing the distance between the experimentally obtained $E(R,\alpha,\tau)$ and the corresponding theoretical value. We need to consider magnon decoherence~\cite{Johansson-CPC-2013} during the tomography operations for the theoretical values $E(R,\alpha,\tau)$. To obtain these theoretical values for fitting the experimental results, we develop a scheme to express the evolution of the decoherence process in a single matrix form, enabling us to conveniently implementing the convex optimization (see Sec.~II.C in Supplementary Materials). The real and imaginary parts of the joint density matrix are shown in Figs.~\ref{fig4}(a) and \ref{fig4}(b), respectively, for the reconstructed Bell state $\rho$.  The fidelity is $\mathcal{F}\equiv\sqrt{\langle \psi_{\rm Bell}\vert\rho\vert\psi_{\rm Bell}\rangle}=0.90\pm0.01$, revealing that the reconstructed state is close to the ideal Bell state. It also indicates that our hybrid magnon-qubit system owns sufficiently high quantum coherence during the process of generating this macroscopic entangled state.

\section{Discussion}
Quantum magnonics emerges from the interplay between quantum information science and spintronics~\cite{Yuan-PR-22}. Also, the quantum regime of a large ferromagnetic object is in itself a cutting-edge research area in both condensed matter and quantum physics. In this work, we have harnessed a magnon-qubit resonant swapping technique in the time domain~\cite{Xu-PRL-2023} to prepare a quantum state of the hybrid system. This allows us to deterministically generate the macroscopic entangled Bell state between the millimeter-size ferromagnetic spin system and the superconducting qubit.
	
However, characterization of the generated quantum states is more challenging compared with the state generation, because it needs more sophisticated quantum-control methods and also requires the system to own higher coherence. As shown in Fig.~\ref{fig3}(a), the sequence for joint quantum tomography is more complicated and takes a time much longer than the generation sequence. Here, we have improved the conventional joint-tomography technique because our scheme no longer requires the qubit to stay in the ground state and be disentangled from the other system. This advancement makes it feasible to efficiently characterize the generated Bell state in our system.

Moreover, our work has not only extended the frontier of macroscopic entangled states to a millimeter-sized ferromagnetic spin system composed of $\sim 10^{19}$ atoms, but also provided opportunities for magnon-based quantum engineering applications such as the quantum transducer~\cite{Hisatomi-PRB-2016,Lachance-APE-2019} and quantum networks~\cite{Lachance-APE-2019,Kimble-Nature-2008,Li-PRXQuantum-2021} that involve different subsystems. It offers an intriguing avenue for further investigations in quantum magnonics.

Note that there are relevant theoretical proposals with the same or similar physical configurations (see, e.g., Refs.~\cite{Kheirabady-Optik-2023,Luo-OL-2021,Liu-PRB-2019}). In Ref.~\cite{Kheirabady-Optik-2023}, the magnon-qubit entanglement dynamics is studied, while Ref.~\cite{Luo-OL-2021} focuses on the magnon entanglement generation in coupled hybrid cavity systems. Also, Ref.~\cite{Liu-PRB-2019} harnesses the same physical configuration but focuses on the magnon blockade. However, our work demonstrates the experimental implementation of the magnon-qubit Bell state. To achieve this maximally entangled state, a tunable superconducting qubit is needed and the state swapping between the qubit and magnons should be implemented in a time much shorter than the decoherence time of the system. Our experiment overcomes these obstacles by using the dressed AT doublet state to tune the qubit frequency and protecting the qubit via a suitably designed cavity. Moreover, we develop a new joint tomography approach to verify the generation of the magnon-qubit Bell state. These are essential distinctions in our work compared to the theoretical approaches.

Finally, it is worth further noting that the magnon Kerr effect plays a negligible role in this study. In Ref.~\cite{Wang-PRB-2016}, we explored this nonlinear effect, which strengthens when reducing the size of the YIG sphere. However, the size of the YIG sphere cannot be too small because the hybrid system needs a sufficiently strong coupling between the YIG sphere and the cavity mode. Thus, we choose a 1 mm-diameter YIG sphere, which has an appreciable magnon Kerr effect when a large number of magnons are excited therein by applying a strong microwave drive to the YIG sphere. Nevertheless, in the present work, only a very weak microwave pulse is applied to the YIG sphere in the state tomography process and, moreover, the whole hybrid system is placed in a chamber at cryogenic temperature. In such a manner, few magnons are generated, giving rise to a negligibly weak magnon Kerr effect. This is the reason we do not observe any appreciable magnon Kerr effect in the present experiment.


\section{Methods}
\subsection{The device and measurement setup}
\begin{figure}
	\includegraphics[width=0.49\textwidth]{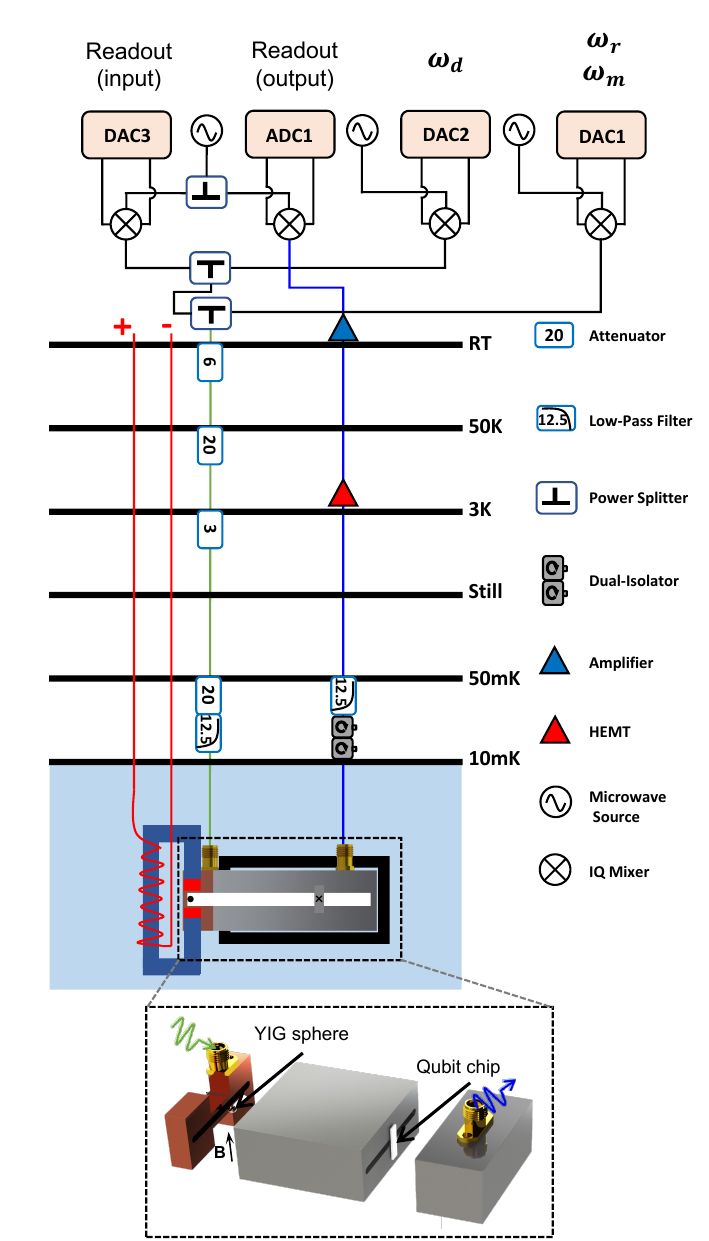}%
	\caption{Schematic diagram of the experimental setup and wiring. It shows the device that is placed in the (blue) mixing chamber of the dilution refrigerator and the measurement system at the top of the figure, which is composed of three DAC boards for signal generation and one ADC board for qubit readout. The DC current for producing the bias magnetic field flows through the (red) superconducting NbTi wire in the electromagnet. Also, the device in the mixing chamber is schematically shown at the bottom of the figure, which includes a YIG sphere and a superconducting qubit embedded in the cooper (left) and aluminium (right) parts of a 3D microwave cavity, respectively.}
	\label{figs1}
\end{figure}

	The device contains three components:~a superconducting-qubit chip, a 3D microwave cavity, and a 1~mm-diameter YIG sphere. In the experiment, the device is placed in the mixing chamber of a dilution refrigerator with a base temperature of 10~mK (see Fig.~\ref{figs1}).

The superconducting qubit is a 3D transmon~\cite{Paik-PRL-2011,Rigetti-PRB-2012} consisting of a single Josephson junction connected to two aluminum pads. These two pads, together with the 3D cavity, provides a large shunt capacitor to reduce the sensitivity of the qubit to the charge noise~\cite{You-PRB-2007, Koch-PRA-2007}. The charging energy of the qubit is measured to be $E_C/h\approx354$~MHz, which is closely related to the anharmonicity of the qubit, i.e., $\eta/2\pi=-E_C/h\approx-354$~MHz. The Josephson coupling energy of the qubit is $E_J/h\approx18.3$~GHz, giving $E_J/E_C\approx51.7$. Room-temperature resistance of the Josephson junction is about $7.2$~K$\Omega$. The qubit lifetime is measured to be $3.65~\rm{\mu s}$.

The microwave cavity is composed of three parts, one small part made of copper and two large parts made of aluminum. The YIG sphere is mounted in the copper part of the cavity, and the qubit chip is mounted in a specifically designed seam between the two aluminum parts of the cavity. Here copper is chosen for allowing the bias magnetic field to penetrate into the copper part of the cavity to adjust the magnon frequency, while aluminum is used to protect the qubit from the magnetic-flux noise. All three parts form a microwave cavity with dimensions $62\times 36\times 2.5$~mm$^3$. The three lowest cavity modes are $\rm{TE_{101}}$, $\rm{TE_{102}}$, and $\rm{TE_{103}}$, with frequencies $4.861$~GHz, $6.388$~GHz, and $8.367$~GHz, respectively. Among the cavity modes, the $\rm{TE_{102}}$ mode serves as the dominant mode to mediate the magnon-qubit interaction via exchanging virtual photons, and the $\rm{TE_{103}}$ mode is used to read out the qubit states~\cite{Lachance-quirion-Science-2020}. The quality factor of the cavity mode $\rm{TE_{102}}$ is about $2.0\times10^4$ at cryogenic temperature.

    In our experiment, we focus on the Kittel mode of magnons in the YIG sphere. Its linewidth is measured to be about $1.46$~MHz.
Here we use an electromagnet and two permanent magnet disks to provide a magnetic field for tuning the magnon frequency. The electromagnet comprises a copper bobbin with about 10000 turns of superconducting coil. Together with the permanent magnet disk, they can supply a magnetic field ranging from 1600 to 2600~Gauss. In Fig.~\ref{figs1}, we schematically show the experimental setup and wiring, which are similar to those in Ref.~\cite{Xu-PRL-2023}. To minimize the thermal noise from reaching the base plate in the chamber at cryogenic temperature ($\sim 10$~mK), we have incorporated a sequence of attenuators in the transmission line at different stages of the dilution refrigerator.

\section{Acknowledgments}
This work is supported by the National Key Research and Development Program of China (Grant No.~2022YFA1405200), the National Natural Science Foundation of China (Grants Nos.~92265202, 11934010, 12174329, 12304558),  the Postdoctoral Fellowship Program of CPSF (No.~GZC20232342) and the Fundamental Research Funds for the Central Universities (No.~2021FZZX001-02). The qubit chip was made in the Micro-Nano Fabrication Center of Zhejiang University.

\section{Author contributions}
J.Q.Y., D.X. and Y.P.W. conceived the experiment, D.X. and X.K.G. designed and performed the experiment, acquired the data and carried out the data analysis under the supervision of J.Q.Y., Y.C.W provided help in acquiring the data, D.X. designed the cavity and qubit chip, and H.K.L. fabricated the qubit chip. All authors contributed to the discussion of the results and the writing of the manuscript.

\section{COMPETING INTERESTS}
The authors declare no competing interests.


\end{document}